 \documentclass[a4paper]{llncs}

 \usepackage{wrapfig}

\usepackage[caption=false]{subfig}
\usepackage{color}
\usepackage[usenames,dvipsnames,svgnames,table]{xcolor}
\usepackage{multirow}
\usepackage{rotating}
\usepackage{graphicx}
\usepackage[linesnumbered, vlined, ruled]{algorithm2e}
\usepackage{booktabs}
\usepackage{amssymb}
\usepackage{amsfonts}
\usepackage{amsmath}
\usepackage{soul}
\usepackage{etex}
\usepackage{extarrows}
\usepackage{relsize}
\usepackage{epstopdf}

\newcommand{\myFontC}[1]{{\fontfamily{pcr}\selectfont #1}} 

\usepackage{url}
 \usepackage{hyperref} 
\hypersetup{ 
    colorlinks,%
    citecolor=black,%
    filecolor=black,%
    linkcolor=black,%
    urlcolor=black,%
pdftitle={Towards Scalable Visual Exploration of  Very Large RDF Graphs},
pdfauthor={Nikos Bikakis, John Liagouris, George Papastefanatos, Timos Sellis},
pdfsubject={Towards Scalable Visual Exploration of Very Large RDF Graphs},
pdfkeywords=
 {graphVizdb} {graph data} {disk based visualization tool} {RDF graph visualization} {spatial indexing} {visualizing linked data} {partition based graph layout} {interactive visualization} {scalable visualization} {large data visualization}  {Visual analytics} {Semantic Web} {LOD} {RDF visualization} {Data exploration}  {million nodes graphs} {big data} {web of data visualization}
}

\newcommand{\eat}[1]{}

 \newcommand{\stitle}[1]{\vspace{2mm}\noindent\textbf{#1}}
 \newcommand{\aknw}[1]{\small\vspace{2.2mm}\noindent\textbf{#1}	 		
 \setlength{\leftskip}{20pt}
}

\begin{document}

\title{
 Towards Scalable Visual Exploration of   \\ Very Large RDF Graphs
 \thanks{This paper appears in 12th Extended Semantic Web Conference (ESWC 2015).}
}

 \author{
Nikos Bikakis$^{1,2}$ \and \hspace{3pt}
John Liagouris$^{3}$ \and \hspace{3pt}
Maria Kromida${^1}$   \and \\
George Papastefanatos$^{2}$ \and  
Timos Sellis$^{4}$ 
}

\institute{
$^{1}$ NTU  Athens, Greece  
$^{2}$ ATHENA Research Center, Greece \\
$^{3}$ ETH Z\"urich, Switzerland  
$^{4}$ RMIT University, Australia
}

\maketitle
\vspace{-4mm}

\begin{abstract}
In this paper, we outline our  work on developing a disk-based
 infrastructure for efficient visualization and graph exploration operations over very   large graphs.
  The proposed platform, called graphVizdb,  
 is based on a novel technique
for indexing and storing the graph.  
Particularly, the graph layout is indexed with a spatial data structure, i.e., an
R-tree, and stored in a database.
In runtime, user operations are translated into 
efficient spatial operations (i.e., window queries) in the backend.
 \end{abstract}
\vspace{-5mm}
 \keywords
 graphVizdb, graph data, disk based visualization tool, RDF graph visualization, spatial, visualizing linked data, partition based graph layout.
 
 \vspace{-2mm}



\section{Introduction}
\label{sec:intro}
 %
%
Data visualisation provides intuitive ways  
for information analysis, 
allowing users to infer correlations
and causalities that are not always possible with
traditional data mining techniques.
The
wide availability of vast amounts of graph-structured data, RDF in the
case of the Data Web, demands for user-friendly methods and tools for data
exploration and knowledge uptake. We consider some core challenges related on
the management and visualization of very large RDF graphs; e.g., the Wikidata RDF graph has more than 300M nodes and edges. 

First, their size   exceeds the capabilities of
memory-based layout techniques and libraries, enforcing disk-based implementations. 
Then, graph rendering is a time consuming process; even drawing a small
part of the graph (containing a few hundreds of nodes) requires considerable time when we assume
real-time systems. The same holds for graph interaction and navigation. Most
operations, such as zoom in/out and move, are not easily implemented to large
dense graphs, as their implementations require redrawing and re-layout
large parts of them.

Related works in the field handle very large graphs through  hierarchical visualization approaches.
Although hierarchical approaches provide fancy visualizations
with low memory requirements, their applicability is heavily
based on the particular characteristics of the input dataset.
In most cases, the hierarchy is constructed by exploiting clustering and partitioning methods \cite{AbelloHK06,Auber04,BastianHJ09,RodriguesTTFL06,TominskiAS09}.
In other works, the hierarchy is defined with hub-based \cite{LinCTWKC13} and density-based \cite{ZinsmaierBDS12} techniques. \cite{ArchambaultMA08} supports ad-hoc hierarchies which are manually defined by the users.
Some of these systems offer a disk-based implementation \cite{AbelloHK06,RodriguesTTFL06,TominskiAS09} whereas others keep the whole graph in main memory \cite{ArchambaultMA08,Auber04,BastianHJ09,LinCTWKC13,ZinsmaierBDS12}.

%
In the context of the Web of Data \cite{DR11,Mazumdar2015,MazumdarPC14,BrunettiAGKN13,VochtDBCVMMW14,AlonenKSH13,bsp14}, there is a large number of tools that visualize RDF graphs (adopting a node-link approach); the most notable ones are
 \mbox{\textit{ZoomRDF}} \cite{ZhangWTY10}, \textit{Fenfire} \cite{HastrupCB08},
 \textit{LODWheel} \cite{SDN11},
\textit{RelFinder} \cite{HeimLS10} and
\textit{LODeX} \cite{BenedettiPB14}.
%
All these tools require the whole graph to be loaded on the UI.
Several tools that follow the same non-scalable approach have also been developed in the field of ontology visualization \cite{DudasZS14,FuN14}.

In contrast to all existing works, we introduce a
generic platform, called \textsf{graphVizdb}, for scalable graph visualization that do not
necessarily depend on the characteristics of the dataset.
The efficiency of the proposed platform is based on a novel technique
for indexing and storing the graph.
The core idea is that in a preprocessing phase, the   graph is drawn, 
using any of the existing graph layout algorithms. 
After drawing the graph, the coordinates assigned to its nodes (with respect to a Euclidean plane)
are indexed with a spatial data structure, i.e., an
R-tree, and stored in a database.
In runtime, while the user is navigating
over the graph, based on the coordinates, specific parts of the graph are retrieved
and send to the user.




 \begin{figure*}[t]
 \subfloat{\includegraphics[scale=1.59]{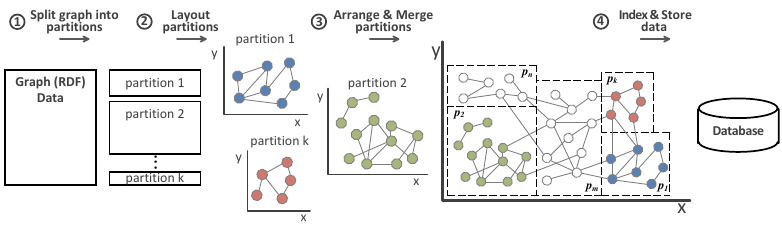}}
 \caption{Preprocessing Overview}
 \label{fig:prep}
 \end{figure*}

\section{Platform Overview}
\label{sec:overview}

The \textsf{graphVizdb} platform is built on top of two main concepts: 
(1) it is based on a ``spatial-oriented" approach for graph visualization,  similar to
approaches followed in browsing maps; and
(2) it adopts a disk-based implementation for supporting interaction with
the graph, i.e., a database backend is used to index and store
graph and visual information.

\stitle{Partition-based graph layout.}
Here we outline the partition-based approach adopted by the \textsf{graphVizdb}
in order to handle very large graph.
 Recall that, for graph layout, the   graph is drawn
once in a preprocessing phase, using any of the existing graph layout algorithms.
However several graph layout algorithms require large amount of memory in order to draw very large graphs.
In order to overcome this problem, our  {partition-based}
approach (outlined in Figure~\ref{fig:prep}) is described next.

$(1)$ Initially, the graph (RDF) data is divided into a set of smaller sub-graphs (i.e., partitions)
using a graph partitioning algorithm.
At the same time, the graph partitioning algorithm tries to minimize the number of edges connecting nodes in different partitions.
$(2)$ Then, using a graph layout algorithm, each of the sub-graph resulted from the graph partitioning, 
is visualized into a Euclidean plane, excluding (i.e., not visualizing) the edges  connecting nodes through different partitions (i.e., crossing edges).
$(3)$ The  visualized partitions are organized and combined  into a ``global"   plane  using a greedy algorithm whose goal is twofold.
First, it
ensures that the distinct sub-graphs do not overlap on the plane,
and at the same time it tries to minimize the total length of the
crossing edges.
$(4)$ Based on the ``global" plane, the coordinates for each node and edge are indexed and stored in the database.

\stitle{Spatial operations for graph exploration.} 
In \textsf{graphVizdb}, most of the user's requests are translated into simple spatial 
operations evaluated over the database.
In this context, \textit{window queries} (i.e., spatial range  queries that retrieve the information  contained with in a specific spatial region) are the core operation for most user's requests. 
The user navigates on the graph by moving
the viewing window. When the window
is moved, its new coordinates with respect to the whole canvas
are tracked on the client side, and a window
query is sent to the server. The query is evaluated on the server using the R-tree indexes.
This way, for each user request, \textsf{graphVizdb} efficiently renders only visible parts of the graph, minimizing in
this way both backend-frontend communication cost as well as
rendering and layout time.
Additionally, more sophisticated operations, e.g., abstraction/enrichment zoom
operations are also implemented using spatial operations.

\stitle{Implementation.}
We have implemented a \textsf{graphVizdb} prototype\footnote{\href{http://graphvizdb.imis.athena-innovation.gr/}{graphvizdb.imis.athena-innovation.gr}} which provides interactive visualization over  large   graphs. 
The prototype offers three main operations: (1) interactive navigation,
(2) multi-level exploration, and (3) keyword search.
We use MySQL  for data storing and indexing, the Jena framework for RDF data handling, Metis\footnote{\href{http://glaros.dtc.umn.edu/gkhome/views/metis}{glaros.dtc.umn.edu/gkhome/views/metis}}  for graph partitioning, and Graphviz\footnote{\href{http://www.graphviz.org}{www.graphviz.org}} for drawing the graph partitions.
In the front-end, we use mxgraph\footnote{\href{http://www.jgraph.com}{www.jgraph.com}},
a client-side JavaScript visualization library.
A video presenting the basic functionality of our
prototype is available at:
\href{https://vimeo.com/117547871}{\myFontC{vimeo.com/117547871}}.

   \bibliographystyle{abbrv}
 \bibliography{main}

  \aknw{Acknowledgement.}   
 {  This research has been co-financed by the European Union (European Social Fund - ESF) and Greek national funds through the Operational Programs ``Education and Lifelong Learning" -  Funding Program: THALIS and ``Competitiveness and Entrepreneurship" (OPCE II) - Funding Program: KRIPIS of the National Strategic Reference Framework (NSRF).  
} 

\end{document}